\documentclass[twocolumn,showpacs,preprintnumbers,amsmath,amssymb,array]{revtex4}
\usepackage{graphicx}
\usepackage{dcolumn}
\usepackage{bm}
\usepackage{amsmath,amsfonts,amssymb,amsthm,verbatim}
\usepackage{floatflt}
\usepackage[ansinew]{inputenc}
\usepackage{multirow}
\usepackage[usenames]{pstcol}
\renewcommand{\i}{\mathrm{i}}
\newcommand{\I}{\mathrm{Ising}}
\newcommand{\id}{\mathbb{I}}
\newcommand{\vect}[1]{\boldsymbol{#1}}
\newcommand{\tabspace}{\vspace{2mm}}
\begin{document}

\title{The Two-Impurity Anderson Model at Quantum Criticality}
\author{David F. Mross$^{1,2}$ and Henrik Johannesson$^2$}
\affiliation{$\mbox{}^1$ Institute of Physics, Bonn University, DE-53115 Bonn, Germany}
\affiliation{$\mbox{}^2$ Department of Physics, G\"oteborg University, SE-412 96 G\"oteborg, Sweden}


\begin{abstract}
We propose a realization of the two-impurity Anderson
model in a double quantum-dot device. When charge transfer
between the dots is suppressed the system exhibits a quantum phase transition,
which is controlled by a surface of non-Fermi liquid
fixed points parameterized by the charge valences of the dots.
Employing conformal field theory techniques, we identify the scaling exponents
that govern transport and thermodynamics close to criticality.
We also determine the dynamical exponents that
set the time scale for the buildup of the non-Fermi liquid state
after the system is suddenly shifted into the critical region, e.g. by
a change of a nearby gate voltage.
\end{abstract}
\pacs{75.20.Hr, 75.30.Hx}
\maketitle

\section{Introduction}

The physics of two interacting quantum impurities embedded in an electron gas is an important and much studied problem. It has bearings on the theory of quantum critical phenomena
\cite{Sachdev} and turns up in a host of other problems, including the study of heavy fermion physics \cite{Coleman}, cluster implementations of dynamical mean-field theory \cite{Ferrero}, and the quest for a solid-state quantum computer \cite{Craig}.

When the impurities carry spin-$1/2$ magnetic moments, the problem becomes that of the two-impurity Kondo model (``TIKM") \cite{Jayprakash}. This model supports two competing energy scales: an Ruderman-Kittel-Kasuya-Yosida (RKKY) interaction $\sim K$ driven by the conduction electrons $-$ which, for $K>0$, tends to lock the magnetic moments into a singlet $-$ and a Kondo temperature $T_K$ below which the electrons try to screen the moments. When the two scales become comparable, there is a crossover between the RKKY-singlet and Kondo screened phases \cite{Sakai}. If the model is endowed with a special kind of electron-hole symmetry, the crossover sharpens into a second-order phase transition, controlled by a non-Fermi liquid fixed point \cite{PhysRevB.52.9528}. However, since the required symmetry is not expected to be found in any known material, the experimental observation of the quantum critical state has been seen as rather unlikely.

In a recent work, Zar\'{a}nd {\em et al.} \cite{zarand:166802} suggested that an experimentally controlled approach to the TIKM quantum critical state may in fact be achieved by using a special type of double-quantum dot device, realized in a gated semiconductor heterostructure. With a design where the two dots are connected to two separate leads, and RKKY-coupled via a magnetic insulator, the quantum critical state was predicted to be insensitive to electron-hole symmetry breaking.

In this paper we address the question what happens if the proposed device is operated in a regime where charge is allowed to fluctuate on the dots (but with charge transfer {\em between} the dots still being suppressed)? An answer to this question is important for validating future experimental tests of the proposal in Ref. \cite{zarand:166802}, as well as for characterizing quantum criticality in the presence of a local perturbation, effectively generated by processes at a higher energy scale. To this end we study the {\em two-impurity Anderson model} (``TIAM") \cite{TIAM}, with the two impurities connected to two separate leads. Applying an extension of the boundary conformal field theory approach (BCFT) that has been used for the TIKM \cite{PhysRevB.52.9528}, we are able to explore the quantum critical properties of this model non-perturbatively. We find that the charge fluctuations generate a {\em surface of unstable fixed points}, connected to that of the TIKM fixed point by two marginal operators, but with thermodynamics and transport properties affected only in system-specific amplitudes. In contrast, the dynamical exponents that govern the approach to the non-Fermi liquid state after the system has suddenly been shifted into the critical region sensitively depend on how much charge is localized on each dot.

\begin{figure}[ht]
\centering
 \includegraphics{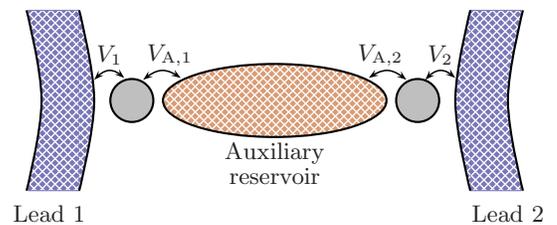}
\centering
\caption{(Color online) The physical system that we study in this article. The different $V$ are tunneling rates. The auxiliary reservoir is operated in the Coulomb blockade regime, where charge transfer between the reservoir and dots (and thus also between the two leads) is strongly suppressed.}\label{fig:device}
\end{figure}

The setup we consider is depicted in Fig. \ref{fig:device}. The two infinite leads are coupled to their respective dots via tunnel junctions, allowing the electrons to hop from [into] lead $i$ into [from] dot $i$ with amplitude $V_i$ $(i=1,2)$. The finite auxiliary electron reservoir (a large quantum dot with a quasi-continuous density of states $\rho$) is coupled to both dots also via tunnel junctions, with tunneling rates $\sim V_{A,i}$.

\normalsize
\normalsize

First consider the case $V_1=V_2=0$. If we tune nearby gate voltages such that each dot has spin 1/2, with charge transfer between the dots and the auxiliary reservoir suppressed by Coulomb blockade \cite{Oreg}, the setup is described by the TIKM, with Hamiltonian
\begin{equation} \label{TIKM}
H_\text{TIKM}=H_\text{kin}+\sum_{i=1,2} J_{A,i}\vect{\sigma}_1\cdot\vect{\sigma}_A(\vect{R}_i)+K\vect{\sigma}_1\cdot\vect{\sigma}_2, \end{equation}
where $J_{A,i} \sim V_{A,i}^2/U_A\ (i\!=\!1,2)$, and $U_A$ being the Coulomb blockade energy of the reservoir.
Here $\boldsymbol{\sigma}_i = d_{i\alpha}^{\dagger} \boldsymbol{\tau}_{\alpha \beta} d_{i\beta}$ and $\boldsymbol{\sigma}_A(\vect{x}) = c^{\dagger}_\alpha(\vect{x}) \boldsymbol{\tau}_{\alpha \beta} c_{\beta}(\vect{x})$, where $d^\dagger_{i\alpha}$ creates an electron with spin $\alpha$ on dot $i$, $c_{\alpha}^\dagger$ creates an electron with spin $\alpha$ in the auxiliary reservoir, and $\vect{R}_1$ and $\vect{R}_2$ denote the locations of the two dots. $\boldsymbol{\tau}_{\alpha \beta}$ are elements of the vector of Pauli matrices, and all repeated spin indices are summed over.
$H_\text{kin}$ is the kinetic energy of the electrons in the auxiliary reservoir. The RKKY interaction with coupling $K \sim J_{A,1}J_{A,2} \cos\varphi /D$ is generated in second order in the Kondo couplings $J_{A,i}$, with $D$ the band width of the reservoir and where $\varphi \sim
\vect{R}_1 - \vect{R}_2$ controls the sign of the interaction. When $K>>T_{A}$, with $T_A$ being the Kondo temperature associated with the auxiliary reservoir, the antiferromagnetic RKKY interaction $K\vect{\sigma}_1\cdot\vect{\sigma}_2$ dominates the direct Kondo interactions \cite{Craig,Vavilov}, and we are left with only the first and last term in the Hamiltonian in Eq. (\ref{TIKM}). Here $T_A=\min (T_{A,1},T_{A,2})$, with $T_{A,i}\sim D \exp(-1/2\rho J_{A,i})$. If we now turn on $V_{1,2}$, charge may fluctuate between each of the dots and their respective leads, {\em but not between the two dots}. By this construction the system can reach a quantum critical phase, protected by the suppression of charge transfer between the two leads \cite{zarand:166802}. As we shall see, this phase is \emph{not} fragile against anisotropies in the couplings $V_i$, as long as charge transfer between the leads is suppressed (which is achieved via Coulomb blockade, as described above). This is an important feature of the proposed device, as fine-tuning of the couplings is not necessary for accessing the critical state.
With the condition that $K>>T_A$, the Hamiltonian modeling the two dots and the electrons in the leads is that of the two-impurity Anderson model,
\begin{equation}H_\text{TIAM}=K\vect{\sigma}_1\cdot\vect{\sigma}_2+H_1+H_2,\label{eqn:TIAM}\end{equation}
where
\begin{eqnarray} H_i & = &H_{\text{kin},i} + \left(V_i c_{i\alpha}^{\dagger}(\vect{R}_i) d_{i\alpha} + h.c.\right) \nonumber \\
& + & U_i(d_{i \uparrow}^{\dagger}d_{i \uparrow})(d_{i \downarrow}^{\dagger}d_{i \downarrow}) + \epsilon_{i}d_{i\alpha}^{\dagger} d_{i\alpha} , \ \ i=1,2. \label{eqn:SIAM} \end{eqnarray}
Here $H_{\text{kin},i}$ is the kinetic energy for the electrons in lead $i$ and $c_{i\alpha}^\dagger$ creates an electron with spin $\alpha$ in lead $i$. $U_i$ is the corresponding charging energy and $\epsilon_{i}$ is the chemical potential that is measured from the common Fermi level of the leads.
If $|\epsilon_i|$ is of the same order of magnitude as the level width $\Gamma_i \sim V_i^2$, its singly occupied subspaces are hybridized with the doubly and unoccupied subspaces (``mixed valence regime''). At the electron-hole symmetric point $\epsilon_i=-U_i/2$ (center of the Coulomb blockade valley of dot $i$) the net charge on the impurity is $n_i=\langle d^\dagger_{i\alpha}d_{i\alpha}\rangle=1$, while $n_i\leq 1$ for larger values of $\epsilon_i$. The $n_i>1$ case need not be considered separately as it is related to $n_i<1$ via an electron-hole transformation.

\section{Identifying the critical theory}

As was first realized by Affleck and Ludwig \cite{NucPhysB360}, the critical properties intrinsic to a quantum impurity problem are encoded by a boundary conformal field theory (BCFT). More precisely, by representing the impurity-electron interaction at criticality by a conformally invariant boundary condition, the electrons can be described by a BCFT. The method is well-documented in the literature \cite{AffleckReview}, and here we only sketch the basics so as to lay down notation and conventions.

Any application of BCFT requires that the electrons are effectively confined to a one-dimensional (1D) geometry. In the present case, the point-like interaction of Eq. (\ref{eqn:SIAM}) affects only the $l=0$ components of the angular-momentum decomposition of the electron fields in each lead. 
The system can thus be described by a trivial non-interacting two-dimensional or three-dimensional theory (involving all higher angular-momentum states) and an interacting 1D theory (involving only the $l=0$ states), where the interesting physics happens and on which we focus exclusively. With this, the first step in the BCFT approach is to set up a framework where the impurity-electron interaction can conveniently be traded for a boundary condition. To this end one may exploit a well-known result from conformal field theory that connects the energy levels of the BCFT on a finite strip (with complex coordinates
$\{ w= u + iv \ | \ -\infty < u < \infty ,\ 0 \leqslant v \leqslant \ell \}$) to the (boundary) scaling dimensions of operators in the upper complex half-plane $\{z = {\tau} + ix \ | \ x \geqslant 0 \}$. (Here $\tau$ is the Euclidean time and $x$ is the spatial coordinate of the 1D model, with $x=0$ being the location of the impurities \cite{footnote2}.) By imposing a conformally invariant boundary condition, call it $A$, at the edges $v=0$ and $v=\ell $ of the strip, and then mapping it onto the semi-infinite plane using the conformal transformation $z = \exp (\pi w/\ell)$ (implying boundary condition $A$
at $x =0$), one obtains a connection between the finite-size energy spectrum on the strip and the
{\em boundary scaling dimensions} in the semi-infinite plane. With $E_0$ as the ground-state energy, one has \begin{equation}  \label{CardyFormula}
E_n = E_0 + \frac{\pi \Delta_n } {\ell}, \end{equation} where $\{ E_n \}$ is the spectrum of excited energy levels in $0 \leqslant v
\leqslant \ell$, and $ \{ \Delta_n \} $ is the spectrum of boundary scaling dimensions. The boundary scaling dimensions $ \Delta_n $ control the asymptotic behavior of the autocorrelation functions close to the boundary,
\begin{multline} \label{Auto}
\langle \varphi_n(\tau,x) \varphi_n(0,x) \rangle - \langle \varphi_n(\tau,x) \rangle
\langle \varphi_n(0,x) \rangle \ \sim \ \tau^{-2\Delta_{n}} \, \\
 |\tau| >> |x|,
\end{multline}
where $\varphi_n$ is a conformal field with the property that the state $\varphi_n |0\rangle$ has energy $E_n$, with $|0\rangle$ being the ground state on the strip.
The autocorrelation functions in turn determine how the presence of the boundary {\em (alias} the impurity-electron interaction) influences the low-temperature thermodynamics and transport in its neighborhood
(treating Euclidean time as an inverse temperature). The problem has thus been reduced to determining the finite-size
spectrum, Eq. (\ref{CardyFormula}), of the theory. First, however, the specific boundary condition $A$ that represents the electron-impurity interaction has to be nailed down. This is a nontrivial task, and we first review how to carry it out in the somewhat simpler case of the TIKM \cite{PhysRevB.52.9528} (which corresponds to the integer-valence limit $U_i=-2\epsilon_i\rightarrow\infty \, (i=1,2)$ of the TIAM).

The two separate channels of 1D electron fields (representing the conduction electrons in leads 1 and 2) each have a U(1)-charge and an SU(2)-spin symmetry. Using non-Abelian bosonization, the fermionic theory can then be rewritten exactly in terms of two U(1)-charge bosons and two SU(2)$_1$ spin bosons, where $k=1$ is the level of the $SU(2)_k$ algebra under which the bosons transform \cite{R}. To incorporate the fact that the Kondo interaction couples the two spin bosons and, therefore, breaks the SU(2)$_1$$\otimes$ SU(2)$_1$ symmetry down to the diagonal subgroup (``total spin"), the spin bosons are expressed in terms of a single SU(2)$_2$ spin boson and $-$ as required by the coset construction \cite{R} $-$ an Ising field. The states in any conformal field theory organize into {\em conformal towers}, corresponding to the irreducible representations of the symmetry group(s) of the theory. In the case of the TIKM (as well as the TIAM) the conformal towers are labeled by the quantum numbers $Q_{1,2}\in\mathbb{Z}$ (net charge in respective lead with respect to the filled Fermi sea), $j\in\{0,1/2,1\}$ (total SU(2)$_2$ spin), and $\lambda \in\{0,1/16,1/2\}$ (corresponding to the Ising fields $\id,\sigma,\epsilon$). The identification of spin conformal towers is shown in Table \ref{tab:coset}.
\begin{table}[ht]\centering
\caption{Identification of spin conformal towers from the coset construction.}\label{tab:coset}
\tabspace
\renewcommand\arraystretch{1.1}
\begin{tabular}[c]{cc}
$\text{SU(2)}_1\otimes \text{SU(2)}_1$&$\text{SU(2)}_2\otimes \I$\\
\hline
$(0)_1\otimes(0)_1$&$\left[(0)_2\otimes \id \right]\oplus \left[(1)_2\otimes \epsilon\right]$\\
$(1/2)_1\otimes(0)_1$&$(1/2)_2\otimes \sigma$\\
$(1/2)_1\otimes(1/2)_1$&$\left[(0)_2\otimes \epsilon \right]\oplus\left[ (1)_2\otimes \id\right]$\\
\end{tabular}
\end{table}
\normalsize

All interactions between the bosons (originating from the Kondo- and RKKY-interactions in the electronic theory) are absorbed into a conformally invariant boundary condition so that the boson dynamics is governed by a free Hamiltonian. The boundary condition is encoded in a set of gluing conditions for combining the conformal towers of the effective (conformal field) theory. The gluing conditions in turn constrain the finite-size energy spectrum from which the dimensions of the scaling operators $-$ and thus the critical scaling of observables $-$ are extracted. In the present case, it has been shown that the particular gluing conditions that describe the impurity-electron interaction at the critical point $-$ which is known to occur for some intermediate value of the RKKY coupling, $K \sim T_K = \mbox{min}\{T_1, T_2\}$, with $T_i \sim (\Gamma_i U_i)^{1/2}\exp(-2\Gamma_i/\pi|\epsilon_{i}|)$ $-$ are obtained from the non-interacting (free-fermion) ones (Table \ref{freeconditions}) via {\em fu-}
\begin{table}[ht]\centering\caption{Gluing conditions for free fermions}\label{freeconditions}
\tabspace
\renewcommand\arraystretch{1.1}
\begin{tabular}[c]{|c|c|c|c|}
\hline
$Q_1$&$Q_2$&$j$&Ising\\
\hline\hline
0&0&0&$\id$\\
\hline\hline
1&0&1/2&$\sigma$\\
\hline
0&1&1/2&$\sigma$\\
\hline\hline
0&0&$1$&$\epsilon$\\ \hline
1&1&1&$\id$\\
\hline
1&1&0&$\epsilon$\\
\hline
\end{tabular}
\end{table}
\normalsize

\noindent {\em sion} with the Ising field $\sigma$ \cite{footnoteHYPOTHESIS}.
This means that an Ising conformal tower gets replaced according to the operator product expansion (OPE) of $\sigma$ with the primary field corresponding to that tower, i.e.
$\sigma\times\sigma=\id+\epsilon$ and $\epsilon \times\sigma=\id\times\sigma=\sigma$.
Since we have a free bosonic theory, the energy spectrum can then be read off immediately as
\begin{equation}E=E_0+\frac{\pi}{4\ell}\left[\left(Q_1\right)^2\!+\!\left(Q_2\right)^2\!+\!j\left(j\!+\!1\right)+\!4\lambda \right],\label{eqn:TIKMenergy}\end{equation}
with the quantum numbers $Q_{1}, Q_{2}, j,$ and
$\lambda$ constrained by the gluing conditions obtained from the ones in Table \ref{freeconditions} by fusion with $\sigma$ (see Table \ref{tab:nontriv}).

The BCFT approach for the TIKM $-$ as reviewed above $-$ can easily be adapted to the TIAM, Eq. (\ref{eqn:TIAM}).
To take into account the fact that, unlike a Kondo impurity, an Anderson impurity (still in the $n_i\!=\!1$ limit) carries charge, we redefine the charges as $Q_i\!\rightarrow \!Q_i \!-\!1$. To keep the correct physics, the gluing conditions must be modified accordingly, by exchanging the $Q_i=1$ and $Q_i=0$ conformal towers (corresponding to a $\pi/2$ phase-shift of the electrons). These replacements are formally equivalent to fusion with the $j_i=1/2$ conformal towers or with the Ising field $\epsilon$ in the coset picture. As pointed out in Ref. \cite{PhysRevB.52.9528}, the critical point is symmetric with respect to $\pi/2$ phase-shifts, which is formally reflected in the OPE $\sigma\times\epsilon=\sigma$. Thus the fusion that gives rise to the proper boundary conditions is not affected by our redefinition of charge.
The operator content at the critical point is obtained via {\em double fusion} \cite{footnote4} with $\sigma\times\epsilon=\sigma$, as in Ref. [\onlinecite{PhysRevB.52.9528}], and one finds five operators with dimension $\Delta<2$ (Table \ref{tab:ops}) that are allowed by conservation of charge and total spin.
\begin{center}
\begin{table}[ht]
\caption{Operator content $(\Delta < 2)$ of the TIKM at criticality. $\boldsymbol{\phi}$ is the $j=1$ spin-boson field. As we will see, the same list of operators and corresponding dimensions describes the critical TIAM.}\label{tab:ops}
\tabspace
\renewcommand\arraystretch{1.1}
\begin{tabular}[c]{c|c}
\hline
Operator&$\Delta$\\
\hline
$\epsilon$&1/2\\
$J^c_{i}\sim\psi^\dagger_{i}\psi_i$&1\\
$J^c_{i}\epsilon\sim\psi^\dagger_{i}\psi_i\epsilon$&3/2\\
$\mathbf{J}^s_{-1}\cdot\boldsymbol{\phi}$&3/2\\
$L_{-1}\epsilon$&3/2\\
\hline
\end{tabular}
\end{table}
\end{center}
\normalsize

In the Kondo limit, where particle-hole symmetry is present and when there is parity symmetry ($\psi_1,d_1\leftrightarrow \psi_2,d_2$), only the first and last operators are allowed \cite{PhysRevB.52.9528}. The operator $\epsilon$ is the only relevant one and comes with a scaling field that measures the distance of the RKKY coupling $K$ to its critical value. Breaking parity symmetry will only allow the \emph{irrelevant} operator $\mathbf{J}^s_{-1}\cdot\boldsymbol{\phi}$, which does not drive the system away from criticality. If we leave the Kondo limit and allow charges to fluctuate between the dots and the leads, i.e., we have $n_i\neq 1$, particle-hole symmetry is no longer present and the second and third operators in Table \ref{tab:ops} become allowed. None of these operators is relevant ($\Delta<1$) and, therefore, criticality is \emph{not} destroyed. The exactly marginal charge currents $J_i^c$ extend the TIKM critical point to a critical surface parameterized by the impurity valences $n_i$. The values of these are determined by the Friedel-Langreth sum rule, which relates the charge $n_i$ of a local perturbation to the electronic scattering phase shift, $\delta_{Fi} = \pi n_i/2$, at the Fermi level. The finite-size energy spectrum is \begin{equation}E=E_0+\frac{\pi}{4\ell}\left[\sum_{i=1,2} (Q_i -n_i)^2+j(j+1)+4\lambda \right],\end{equation}\label{eqn:TIAMenergy}
where the phase shifts are imparted by the exactly marginal charge currents $J_i^c$, and the quantum numbers $Q_1, Q_2, j$, and $\lambda$ are constrained by the same gluing conditions as in the case of the TIKM (Table \ref{tab:nontriv}).
\begin{table}[ht]\centering\caption{Gluing conditions at criticality}\label{tab:nontriv}
\tabspace
\renewcommand\arraystretch{1.1}
\begin{tabular}[c]{|c|c|c|c|}
\hline
$Q_1$&$Q_2$&$j$&Ising\\
\hline\hline
0&0&0&$\sigma$\\
\hline\hline
1&0&1/2&$\id$\\
\hline
0&1&1/2&$\id$\\
\hline
1&0&1/2&$\epsilon$\\
\hline
0&1&1/2&$\epsilon$\\
\hline\hline
0&0&$1$&$\sigma$\\ \hline
1&1&1&$\sigma$\\
\hline
1&1&0&$\sigma$\\
\hline
\end{tabular}
\end{table}
\normalsize

The impurity valences $n_i$ contribute to the spectrum as
$E(n_i)=(\pi/4\ell)\sum_{i=1,2}\left(n_i^2-2n_iQ_i\right)$.
The quadratic term is identical for all states, regardless of their quantum numbers and just offsets the ground state energy $E_0$. The linear part shifts the energy of states with different charges relative to each other. This shift is of the same order as the energy gaps between the different states and changes the spectrum qualitatively. The lowest energy excitation in the TIKM limit for instance becomes degenerate with the second lowest at $n_{1,2}=3/4$ and has higher energy for even lower values.

\section{Thermodynamics and transport}

In contrast to the finite-size energy spectrum, the scaling dimensions of the operators which determine thermodynamics and transport do \emph{not} depend on the impurity valences. This is because the scaling dimensions of these operators related to the energy levels of the finite system with the {\em same} boundary condition on either side of the strip, and are, hence, obtained via {\em double fusion} \cite{footnote4}. As explained in Ref. \cite{PhysRevB.68.075112}, the replacement $Q_i\rightarrow Q_i-n_i$ at the lower boundary corresponds to the replacement $Q_i\rightarrow Q_i+n_i$ at the upper boundary, thus the contributions from both boundaries cancel each other and there is no $n_i$-dependence. The list of operators and corresponding dimensions as given in Table \ref{tab:ops} is therefore valid for the TIAM with arbitrary values of the impurity valences, $0<n_i\leq1 \, (i=1,2)$.

The low-temperature impurity thermodynamics, as well as transport properties due to the presence of the impurity, are determined by the dimension of the leading irrelevant operators, which for the present case are listed as the last three operators in Table \ref{tab:ops}. The first of these, $J^c_i\epsilon$, is a novel composite operator that appears in this model. In the parity symmetric case it governs both the impurity thermodynamics and finite temperature impurity corrections to electron-electron correlation functions. The next operator in the list, $\boldsymbol{J}^s_{-1}\cdot\boldsymbol{\phi}$, is only allowed if parity ($\psi_1,d_1\leftrightarrow \psi_2,d_2$) is broken. In this case it produces the same scaling as known from the overscreened two-channel Kondo model (TCKM) \cite{NucPhysB360}. The last operator in the table, $L_{-1}\epsilon$, is a {\em Virasoro first descendant} and does not give any contribution to finite-temperature properties, contrary to what is claimed in Ref. \onlinecite{zarand:166802} (in context of the TIKM) \cite{footnote3}.
The presence of the composite $J^c_i\epsilon$ operator, and $-$ in the case of broken parity symmetry $-$ the operator $\boldsymbol{J}^s_{-1}\cdot\boldsymbol{\phi}$, produces non-analytic scaling of observables, signaling a non-Fermi liquid ground state. The scenario is similar to that of the TCKM, which is well covered in the literature. An in-depth discussion in the same language as we use here can be found in Ref. \cite{NucPhysB360}. In particular, the first correction to the zero-temperature conductance $G_{0}$ across each dot, generated by $J^c_{i}\epsilon$, scales as
\begin{equation}\label{G} \delta G(T) \sim \sqrt{T/T_K}, \end{equation}
and the contribution to the specific heat is
\begin{equation}\label{C} \delta C(T)\sim T \ln (T_K/T). \end{equation}
The impurity susceptibility is not affected by $J_i^c\epsilon$, since these operators do not contain any of the SU(2)$_2$ spin-boson fields that could couple to a magnetic field. However, when parity symmetry is broken, e.g. by tuning the tunneling rates such that $V_1 \neq V_2$, the operator $\boldsymbol{J}^s_{-1}\cdot\boldsymbol{\phi}$ comes into play, giving rise to a logarithmically divergent term in the susceptibility,
\begin{equation}\label{chi} \delta\chi(T)\sim\ln (T/T_K), \end{equation}
thus replacing the ``screened'' behavior, $\chi(T) = \text{const.} + {\cal O}(T)$, in the parity-symmetric case.

Let us conclude this section by pointing out that while the impurity valences $n_i$ do not affect the scaling to any order, they are expected to appear in the non-universal amplitudes that multiply the expressions in Eqs. (\ref{G}) - (\ref{chi}). To pinpoint the exact dependence of these amplitudes on $n_i$, however, requires methods beyond a BCFT approach \cite{footnoteADD}.

\section{Buildup of the critical ground state: Asymptotic scaling}

\begin{figure}[ht]
\includegraphics{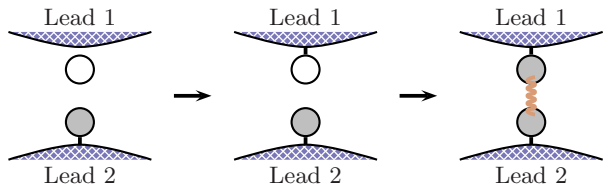}
\vspace{1mm}
\caption{(Color online) One of the physical processes that we consider in this section (case C$_1$ in Table V). In the first step there is a SIAM and a decoupled channel of free electrons. In the second step, the coupling $V_1$ is suddenly turned on, allowing charge to fluctuate between the upper lead and the upper dot. In the last step the system has relaxed to its critical ground state due to the RKKY interaction between the two dots.}\label{fig:remove}
\end{figure}

We now shift focus, and consider a situation where one or several of the tunnel junctions in the device are ``pinched off" (by a change of the gate voltages that control the corresponding junctions), implying that the electrons in the two leads become dynamically decoupled. At a later time $-$ after the severed system has relaxed to equilibrium $-$ the device is reconnected by a sudden ``switch on'' of the same tunnel junctions (again by a change of gate voltages). This poses a question that is a variant of the Fermi edge singularity problem: ``How much time does the system need for rebuilding the non-Fermi liquid critical state after the sudden perturbation?''. To answer this question we consider a setup initially described by two decoupled Fermi liquids, the exact nature of which depends on which tunnel junctions are considered. Then, at some time $t=t_0$ we suddenly reconnect the system by turning on these tunnel junctions. If the parameters of the setup are properly tuned, the system will evolve into the quantum critical phase associated with the TIAM non-Fermi liquid ground state (Fig. \ref{fig:remove}).
\normalsize

We now systematically consider the different scenarios that can emerge, corresponding to turning off (and on) the various (combinations of) couplings (see Table\ref{processes}).
\begin{table}[ht]\centering
\caption{List of ``buildup" processes. The boldface crosses denote couplings that are first turned off, and then $-$ after the severed system has equilibrated $-$
are suddenly turned on again. Taking $1\leftrightarrow 2$ completes the list to all possible combinations. The crosses in normal font mark couplings that may or may not be turned off and on without changing the character of the respective process.}.\label{processes}
\newline
\small
\renewcommand\arraystretch{1.2}
\begin{tabular}[c]{|c|c|c|c|c|l|}
\hline
&$V_1$&$V_2$&$V_A^{1}$&$V_A^{2}$&initial lead configurations\\
\hline
\multirow{2}{*}{A}&\multirow{2}{*}{\bf{X}}&\multirow{2}{*}{\bf{X}}&\multirow{2}{*}{\textcolor[rgb]{0.50,0.50,0.50}{X}}&\multirow{2}{*}{\textcolor[rgb]{0.50,0.50,0.50}{X}}&1. free electrons\\&&&&& 2. free electrons\\\hline
\multirow{2}{*}{B}&\multirow{2}{*}{\bf{}}&\multirow{2}{*}{}&\multirow{2}{*}{\bf{X}}&\multirow{2}{*}{\textcolor[rgb]{0.50,0.50,0.50}{X}}&1. SIAM\\&&&&& 2. SIAM\\\hline
\multirow{2}{*}{C$_1$}&\multirow{2}{*}{\bf{X}}&\multirow{2}{*}{}&\multirow{2}{*}{\textcolor[rgb]{0.50,0.50,0.50}{X}}&\multirow{2}{*}{\textcolor[rgb]{0.50,0.50,0.50}{X}}&1. free electrons\\&&&&& 2. SIAM\\\hline
\multirow{2}{*}{C$_2$}&\multirow{2}{*}{\bf{X}}&\multirow{2}{*}{}&\multirow{2}{*}{\textcolor[rgb]{0.50,0.50,0.50}{}}&\multirow{2}{*}{\textcolor[rgb]{0.50,0.50,0.50}{}}&1. free electrons\\&&&&& 2. free electrons\\\hline
\end{tabular}
\end{table}
\normalsize
Before we proceed with details, let us briefly discuss the various possible processes. Process ``A'' is maybe the most obvious one: If both couplings between the leads and the dots are turned off, the initial state of the conduction electrons is that of a free electron gas. Note that there are actually three different variations of this process, namely with both (decoupled) dots empty, with one of them empty and one of them filled, or with both of them filled. Process ``B'', i.e. turning off one of the dot-reservoir couplings, say $V_{A,1}$, turns the system into a SIAM for lead 1 and its attached dot, while the rest of the system gets described by an anisotropic two-channel SIAM (with the two channels provided by lead 2 and the auxiliary reservoir, respectively). Since the latter model renormalizes to a single-channel SIAM \cite{nozieres} the conduction electrons are effectively described by two decoupled SIAMs, just as if both $V_{A,1}$ and $V_{A,2}$ were turned off. The last two processes in Table V, C$_1$ and C$_2$ $-$ both corresponding to turning off the coupling $V_{1}$ $-$ are a bit more involved and depend on whether the dot $1$ carries charge or not after decoupling from lead $1$. Certainly, the conduction electrons on lead $1$ are described by a free electron-gas in either case. If the dot $1$ carries no charge, then lead $2$ is described by an anisotropic two-channel SIAM which renormalizes to a single-channel SIAM, as before. On the other hand, if the dot is charged, it will form a singlet with dot $2$ since, by the constraint $K \gg T_A$, the RKKY interaction dominates the Kondo interaction. As a consequence, the second lead decouples, turning it into a free electron gas.

\begin{figure}[ht]
\includegraphics{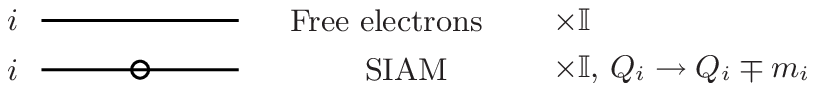}
\caption{(Color online) Notation for the single Fermi-liquid leads ($i=1,2$) and corresponding fusion/redefinitions. Here $\pi m_i/2$ is \emph{defined} as the phase shift that the electrons experience at $x=0$, such that $m_i$ equals the net charge on the dot $\langle d^\dagger_i d_i\rangle$ when it is coupled to the leads (and $m_i=0$ when the coupling is turned off, independent of $\langle d^\dagger_i d_i\rangle$). The sign in front of $m_i$ depends on whether the dot is attached to the lower (+) or upper boundary (-) of the finite strip.}\label{singleglue}
\end{figure}

In the language of BCFT the interactions with the impurities only appear via the boundary condition, and hence the physical process of changing couplings corresponds to changing the boundary conditions. This can formally be described by an appropriate \emph{boundary-changing operator} acting on the system. To make this a bit clearer, we introduce a notation as depicted in Fig. \ref{singleglue} for each of the two possible channels
that may enter into the initial configuration (the severed system at equilibrium): free electrons [isolated lead] and SIAM [lead with an attached dot].
Since the severed system consists of {\em two} leads (with, or without, an attached dot) both must be specified in order to determine the initial configuration. We here note that for a decoupled system, the fusion and charge redefinition that take the free electron gas onto a SIAM can be performed successively. The trivial fusion with the identity operator $\mathbb{I}$ (cf. FIG. 3) reflects the fact that the low-energy physics of the SIAM is that of a local Fermi liquid, hence the gluing conditions on the charge- and spin sectors are the same as for free electrons.

\normalsize

The charge redefinition $Q_i\rightarrow Q_i\mp m_i$ follows as before from the Friedel-Langreth sum rule. A remark is in order concerning $m_i$, which is the net charge on dot $i$ when it is coupled to lead $i$ but decoupled from the other dot. It does not have to be equal to $n_i$, which is the net charge of the same dot at the critical point of the TIAM (with the two dots being RKKY-coupled). In fact, we expect that $n_i\leq m_i$ $(n_i,m_i\in[0,1])$. Since only the singly occupied subspaces participate in the RKKY interactions, the effective RKKY coupling increases with increasing $n_i$. The RKKY coupling is antiferromagnetic ($K>0$) and the spin-spin expectation value at the critical point is negative \cite{PhysRevB.40.324}, thus, larger values of $n_i$ are energetically favorable.
\begin{figure}[ht]
\includegraphics{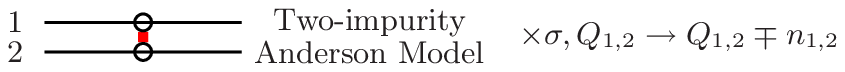}
\caption{(Color online) Notation for coupled leads and corresponding fusion/charge-redefinitions.}\label{doubleglue}
\end{figure}
\normalsize

\begin{figure}[ht]
\centering
\includegraphics{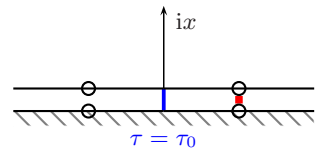}
\caption{(Color online) A process where the initial state consists of two decoupled SIAMs, where the boundary condition is changed at $\tau_0$ to that of the TIAM.}\label{fig:process}
\end{figure}

We represent a decoupled system by drawing the two leads parallel to each other, using the notation from FIG. 3. The two leads can be connected only by the RKKY interaction, which we denote by a line connecting the impurities (Fig. \ref{doubleglue}).
Using this notation, we can represent the process of turning on the couplings (lead-dot tunnel junctions and/or the RKKY interaction) at $\tau=\tau_0=\i t_0$ as shown in Fig. \ref{fig:process}, where the initial configuration is chosen here as two SIAMs (entry B in Table V). The quantity that measures how far the system has evolved from its initial state after a time $\Delta \tau$, is given by the Green's function of the (yet undetermined) boundary-changing operator $\mathcal{O}$: \begin{equation}G(\tau)=\langle 0, \text{I} | \mathcal{O}(\tau_0+\Delta \tau)\mathcal{O}^\dagger(\tau_0)|0,\text{I}\rangle\sim(\Delta \tau)^{-2x},\end{equation} where $x$ is the scaling dimension of $\mathcal{O}$ and $|0,\text{I}\rangle$ is the initial ground state of the severed system.

\normalsize

\begin{figure}[ht]
\includegraphics{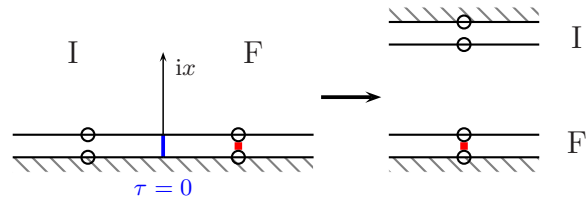}

\caption{(Color online) Mapping to the strip geometry results in a theory with boundary condition I on one side and boundary condition F on the other.}\label{fig:mapping}
\end{figure}

The boundary-changing operator is assumed to be a primary field so that the asymptotic states are ground states of their respective configurations. In this case the Green's function is slaved to the given asymptotic form by conformal invariance. To relate $x$ to the energy spectrum, we map the half plane onto the infinite strip via the conformal transformation $u+\i v= (\ell/\pi) \ln (\tau+\i x)$ (Fig. \ref{fig:mapping}), following the prescription laid out in Ref. [\onlinecite{0305-4470-27-16-007}].

\normalsize

\noindent Under this transformation the propagator takes the form
\begin{multline}
\langle 0,\text{I}\text{I}|\mathcal{O}(u_0+\Delta u)\mathcal{O}^\dagger(u_0)|0,\text{I}\text{I}\rangle =  \, \left(\frac{2\ell}{\pi}\sinh\frac{\pi}{2\ell}\Delta u\right)^{-2x} \\
\stackrel{\Delta u \gg \ell}{\longrightarrow}  \, \, \, \frac{\pi}{\ell}e^{-\frac{\pi}{\ell} x \Delta u} , \label{eqn:dim}
\end{multline}
where $u_0= (\ell/\pi) \ln(\tau_0)$.
Here we denote the ground state of the system on the strip geometry (same boundary condition on both sides) by $|0,\text{I}\text{I}\rangle$.
Alternatively, the propagator can also be evaluated in the strip geometry by inserting a complete set of states. Note that since $\mathcal{O}$ changes the boundary conditions, the eigenstates of the system after applying $\mathcal{O}$ are not the same as the ones before (see Fig. \ref{fig:mapping}). We label states with boundary condition I on the top and boundary condition F on the bottom (see right hand side of Fig. \ref{fig:mapping}) by $|n,\text{I}\text{F}\rangle$. These are the eigenstates after applying $\mathcal{O}$ and a complete basis of them should therefore be inserted,

\begin{align}
&\langle 0,\text{I}\text{I}|\mathcal{O}(u_0+\Delta u)\mathcal{O}^\dagger(u_0)|0,\text{I}\text{I}\rangle\nonumber\\
=&\sum_n\langle 0, \text{I}\text{I}|e^{H\Delta u}\mathcal{O}(u_0)e^{-H \Delta u}|n,\text{I}\text{F}\rangle\langle n,\text{I}\text{F}|\mathcal{O}^\dagger(u_0)|0,\text{I}\text{I}\rangle\nonumber\\
=&\sum_n \left|\langle 0, \text{I}\text{I}|\mathcal{O}(u_0)|n,\text{I}\text{F}\rangle\right|^2 e^{\Delta u\left(E^{\text{I}\text{I}}_0-E^{\text{I}\text{F}}_n\right)}.
\end{align}

In the limit of large $\Delta u$, only the lowest-energy state $|n_0,\text{I}\text{F}\rangle$ contributes and by comparison with Eq. (\ref{eqn:dim}), we get:
\begin{equation}x=\frac{\ell}{\pi}\left(E^{\text{I}\text{F}}_{n_0}-E^{\text{I}\text{I}}_0\right).\label{dimensionenergy}
\end{equation}

So far, no particular boundary-changing operator has been specified. Since the scaling dimension of a boundary-changing operator depends only on the energy of the corresponding state on the strip, with one type of boundary condition at the bottom and another type of boundary condition on the top, the possible boundary-changing operators are determined by the gluing conditions in this geometry.
Using the notation introduced above (cf. Figs. \ref{singleglue} and \ref{doubleglue}), it is a straightforward task to find the correct operators and their scaling dimensions.
The energy is measured with respect to the ground state of the system with boundary condition II, which corresponds to two decoupled SIAMs on each side of the strip.
According to Fig. \ref{singleglue} the gluing conditions are the ones for free fermions (Table \ref{freeconditions}). Moreover, the redefinitions of the charge on either side of the strip cancel: $Q_i\rightarrow Q_i+m_i-m_i=Q_i$. Thus $E^{\text{II}}_0$ is indeed equal to the ground-state energy of the free fermion theory $E_0$.

Since the SIAM connects continuously to the free electron gas via the charge redefinitions $Q_i \mp m_i$ it is sufficient to consider two decoupled SIAMs as the initial state and get the other possibilities by setting the appropriate $m_i$ to zero and keeping track of the correct quantum numbers $Q_1,Q_2$.
For the configuration labeled by IF (Fig. \ref{fig:mapping}, right hand side), the boundary condition is that of two decoupled SIAMs at the top and the one of the TIAM at the bottom of the strip.
By comparison with Figs. \ref{singleglue} and \ref{doubleglue} this translates into fusion with $\sigma$ and the charge redefinitions $Q_1\rightarrow Q_1-n_1+m_1,Q_2\rightarrow Q_2-n_2+m_2$. Defining $\tilde{n}_i\equiv n_i-m_i$ the energy for states with these boundary condition is
\begin{equation}E=E_0+\frac{\pi}{4\ell}\left[\, \sum_{i=1,2} (Q_i -\tilde{n}_i)^2\!+\!j(j+1)\!+\!4\lambda\right].\label{stripenergy}\end{equation}\normalsize
Combining this with Eq. (\ref{dimensionenergy}) immediately yields the exponents for the various processes (see Table \ref{semiconditions}).

\begin{table}[ht]\centering\caption{Exponents governing the buildup of the quantum critical ground state for the various processes. To distinguish between the different possibilities for each process (see discussion following Table \ref{processes}), we have specified the initial net charge on the dots $( \langle d^\dagger_1d_1\rangle,\langle d^\dagger_2 d_2\rangle )$ for each process.}.\label{semiconditions}
\\
\small
\renewcommand\arraystretch{1.45}
\begin{tabular}[c]{c|c|c|}
\cline{2-3}
&Process&$x_i$\\
\cline{2-3}\cline{2-3}
\large 1\normalsize&A(0,0), B($m_1$,$m_2$), C$_1$(0,$m_2$)&$\frac{1}{16}+\frac{1}{4}\tilde{n}_1^2+\frac{1}{4}\tilde{n}_2^2$\\ \cline{2-3}
\large 2 \normalsize&A(1,0), C$_1$(1,$m_2$)&$\frac{7}{16}-\frac{1}{2}\tilde{n}_1+\frac{1}{4}\tilde{n}_1^2+\frac{1}{4}\tilde{n}_2^2$\\ \cline{2-3}
\large 3 \normalsize&A(1,1), C$_2$(1,1)&$\frac{9}{16}-\frac{1}{2}\tilde{n}_1-\frac{1}{2}\tilde{n}_2+\frac{1}{4}\tilde{n}_1^2+\frac{1}{4}\tilde{n}_2^2$\\
\cline{2-3}
\end{tabular}
\end{table}
\normalsize

\begin{figure}[ht]
\centering
\includegraphics{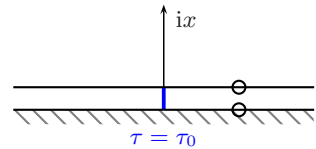}
\caption{(Color online) Two decoupled channels of electrons are turned into two decoupled SIAMs at $\tau_0$.}\label{fig:decouple}
\end{figure}

\noindent
\newline
\newline
The Green's function of the three different boundary changing operators
\begin{equation*}G_{ij}(t)=\langle 0|\mathcal{O}_i(t)\mathcal{O}_j^\dagger(0)|0\rangle\sim t^{-2x_i}\delta_{ij}, \ \ \ i=1,2, 3\label{eqn:boundarygreen}\end{equation*}
measures the overlap between the initially perturbed state and the state to which the system has evolved at time $t$, with $x_i$ listed in Tab (VI).
It is interesting to compare these exponents with the ones for two decoupled SIAMs, i.e., the process depicted in Fig. \ref{fig:decouple}, where the $m_i$ of the two SIAMs take the role of the $\tilde{n}_i$ for the TIAM. The analogy is most obvious in the cases A and C$_2$ when $\tilde{n}_i = n_i$, i.e., when the initial state consists of two channels of free electrons, (as in Fig. \ref{fig:decouple}).

\normalsize

\noindent Since the two SIAMs are completely independent, the exponents for two such models \cite{Menge} can simply be added and we find that our exponents in Table VI (upon identifying $m_i$ and $\tilde{n}_i$) differ from those only by constants of $\pm 1/16$, where $1/16$ is the scaling dimension of the Ising field $\sigma$. The fact that the valences guide the respective processes in this very similar way is notable, considering that the TIAM critical ground state is of non-Fermi liquid type.

\section{Summary}

In conclusion, we have proposed a realization of the two-impurity Anderson model in a double-quantum dot device with separated leads. We have presented an exact BCFT description of the system at quantum criticality and used it to calculate several physical properties. We have found a new composite operator that governs the low-temperature behavior, and explained how non-Fermi liquid scaling behavior which is not present in the original TIKM \cite{PhysRevB.52.9528} emerges as a consequence of certain (irrelevant) symmetry breakings. Whereas the energy spectrum explicitly and qualitatively depends on the charge valences of the dots, thermodynamics and transport properties are only affected in non-universal amplitudes. Instead, the charge valences enter in the dynamical exponents that set the time scale for buildup of the non-Fermi liquid quantum critical state.
\newpage

{\bf Acknowledgments.} \ We thank C.-H. Chung and H. R. Krishna-murthy for valuable comments and suggestions. We also thank the KITP at UCSB for hospitality during the completion of this work. This research was supported in part by the National Science Foundation under Grant No. PHY05-51164, and by the Swedish Research Council under Grant No. VR-2005-3942.


\begin{thebibliography}{99}
\bibitem{footnote2}
Due to the special form (point-like hybridization and coupling of the subsystems via spins only) of the interactions in Eq. (\ref{eqn:SIAM}), we can use the trivial redefinitions $c_{\i\alpha}(\vect{x})\rightarrow c_{i\alpha}(\vect{x}-\vect{R}_i)$ to formally obtain $\vect{x}=0$ as the impurity position in both of the leads.
\bibitem{footnoteHYPOTHESIS}
The statement that the boundary condition representing the impurity-electron
interaction in the TIKM is obtained by fusion with the Ising field $\sigma$ is strictly speaking
a hypothesis within BCFT, and as such must be tested against independent results. This was
done in Ref. [\onlinecite{PhysRevB.52.9528}] where it was shown that fusion with $\sigma$
produces a finite-size spectrum in excellent agreement with numerical results for the TIKM.
\bibitem{footnote3}
The reason that $L_{-1}\epsilon$ does not contribute to the finite-temperature properties lies in the symmetry of the finite temperature geometry. The potential contributions of $L_{-1}\epsilon$ to the partition function are of the form
$\int_0^\beta d\tau \left\langle L_{-1}\epsilon(\tau,0)\ldots\right\rangle_T, $
where the three dots in the correlator represent any other set of operators not depending on $\tau$ and the subscript $T$ means that the correlator is evaluated in the finite-temperature geometry, i.e. a cylinder where $\tau=0$ and $\tau=\beta$ are identified. Using $L_{-1}\epsilon(\tau)=\partial_\tau \epsilon(\tau)$ such terms are seen to vanish:
$\int_0^\beta d\tau \partial_\tau \langle\epsilon(\tau,0)\ldots\rangle=\langle\epsilon(\beta)\ldots\rangle-\langle\epsilon(0)\ldots\rangle=0. $
\end{thebibliography}

\end{document}